\documentclass{article}

\usepackage[final]{neurips_2023}

\usepackage[utf8]{inputenc} 
\usepackage[T1]{fontenc}    
\usepackage{hyperref}       
\usepackage{url}            
\usepackage{booktabs}       
\usepackage{amsfonts}       
\usepackage{nicefrac}       
\usepackage{microtype}      
\usepackage{xcolor}         

\usepackage{amsmath}
\usepackage{amssymb}
\usepackage{graphicx}
\newcommand\independent{\protect\mathpalette{\protect\independenT}{\perp}}
\def\independenT#1#2{\mathrel{\rlap{$#1#2$}\mkern2mu{#1#2}}}

\usepackage{algorithm}
\usepackage{algpseudocode}

\title{Multi-View Interactive Collaborative Filtering}

\author{%
  Maria R. Lentini, Umashanger Thayasivam \\
  College of Science and Mathematics \\
  Rowan University\\
  Glassboro, NJ 08028 \\
  \texttt{lentin26@students.rowan.edu} \\
}

\begin{document}

\maketitle

\begin{abstract}
     In many scenarios, recommender system user interaction data such as clicks or ratings is sparse, and item turnover rates (e.g., new articles, job postings) high. Given this, the integration of contextual ``side" information in addition to user-item ratings is highly desirable. Whilst there are algorithms that can handle both rating and contextual data simultaneously, these algorithms are typically limited to making only in-sample recommendations, suffer from the curse of dimensionality, and do not incorporate multi-armed bandit (MAB) policies for long-term cumulative reward optimization. We propose multi-view interactive topic regression (MV-ICTR) a novel partially online latent factor recommender algorithm that incorporates both rating and contextual information to model item-specific feature dependencies and users' personal preferences simultaneously, with multi-armed bandit policies for continued online personalization. The result is significantly increased performance on datasets with high percentages of cold-start users and items.
\end{abstract}

\section{Introduction}
Traditional collaborative filtering (CF) algorithms such as alternating least squares (ALS) \cite{zhou2008large}, non-negative matrix factorization (NMF), neural collaborative filtering (NCF) \cite{he2017neural} or Bayesian personalized ranking (BPR) \cite{rendle2012bpr} have no mechanism for including contextual information (contextual variables are defined as covariates which describe users and items individually or simultaneously) and are unable to make out-of-sample (i.e., cold-start) predictions. Contextual bandits (CB) \cite{li2010contextual} and factorization machines (FM) \cite{rendle2012factorization} can learn on combined contextual and rating data and make out-of-matrix predictions by modifying the input data accordingly. However, when there are large numbers of distinct users or items, or when there are categorical variables, documents composed of words, or feature ontologies, the dimensionality of the design matrix can blow up. CB scales poorly with dimensionality, and while FM scales linearly, both models suffer from the curse of dimensionality. Furthermore, these algorithms cannot be combined with multi-armed bandit (MAB) policies such as Thompson Sampling (TS) or Upper Confidence Bound (UCB).

There are several state-of-the-art models which address both the curse of dimensionality and the cold-start problem simultaneously. Probabilistic matrix factorization (PMF) \cite{mnih2007probabilistic} reduces the dimension of the design matrix and is similar to traditional matrix factorization (MF). Collaborative topic regression (CTR) \cite{wang2011collaborative}, combines PMF \cite{mnih2007probabilistic} with latent topic modeling, and learns on item-specific (but not user-specific) contextual variables. Interactive collaborative filtering \cite{zhao2013interactive} combines PMF with MAB policies such as epsilon-greedy, Thompson Sampling (TS), upper confidence bound (UCB) and GLM-UCB. The probabilistic frameworks mentioned above address the cold-start problem by relying on pre-specified priors. However, using diffuse priors, as these models do, results in poor predictive performance on cold-start events. We therefore propose a model which personalizes priors using user-item feature dependencies to improve cold-start recommendations. 

Interactive collaborative topic regression (ICTR) \cite{wang2018online} is another probabilistic algorithm which explicitly models item dependencies as user preference clusters. The built-in modeling of arm dependencies helps the algorithm learn faster, but otherwise it too depends on diffuse priors for generating cold-start recommendations. The model we propose is similar, but instead of modeling arm dependencies it models user-item feature dependencies. Taking inspiration from BayesMatch (BM) \cite{maurya2017bayesian}, a multi-view probabilistic clustering algorithm, we develop RatingMatch (RM), which clusters positively associated (i.e., via implicit or explicit ratings) user and item features. When combined with PMF and a bandit policy we call the resulting algorithm \textit{multi-interactive collaborative filtering} (MV-ICTR).

Note that our framework, inspired from CTR \cite{wang2011collaborative} does allow for the integration of both the ICTR from \cite{wang2018online} with RatingMatch to leverage the strengths of both models. Note that CTR does not consider rating data in the procedure for learning its topic components, relying on latent Dirichlet analysis (LDA) \cite{blei2003latent}, whereas BayesMatch is formulated to explicitly leverage ratings as the associative bodies (i.e., user and item feature sets grouped in proportion to rating as opposed to words grouped by merit of being contained within the same document).

 MV-ICTR has dimensionality reduction built-in, for improved performance and reduced computational complexity, separates the tasks of cold-start recommendation and online personalization and thereby improves the short and long-term user experience. 

\section{Background}
\subsection{Collaborative topic regression}
 We begin with a quick over view of collaborative topic regression (CTR). CTR combines topic modelling and CF \cite{wang2011collaborative} using a PMF framework, where the rating $r_{ij}$ for user $i$ on item $j$ is assumed to come from a normal distribution with mean given by the inner product of latent user and item feature vectors:

 \begin{align}
    r & \sim \mathcal{N}(
        u^T_i v_j, c^{-1}_{ij} 
    ) \\
    \label{reward_distr}
 \end{align}

 where $u_i$, $v_j \in \mathbb{R}^d$ are the user and item latent feature vectors, respectively. The precision parameter $c_{ij}$ gives the confidence for rating $r_{ij}$ and is defined heuristically as:

 \begin{align}
     c_{ij} = 
        \begin{cases} 
          a & r_{ij} = 1 \\
          b & r_{ij} = 0
        \end{cases}
 \end{align}

 with $a > b > 0$, since we are more confident in clicks indicating positive sentiment and less confident in the absence of a click indicating negative feedback. The sampling procedure is as follows:

 \begin{align}
    \epsilon_j & \sim \mathcal{N}(0, \lambda^{-1}_v I_k) \\
    \theta_j & \sim \text{Dirichlet}(\alpha) \\
 \end{align}

 Letting $v_i = \theta_j + \epsilon_j$.

  \begin{align}
  \begin{split}
    u_i & \sim \mathcal{N}(0, \lambda^{-1}_u I_K) \\
    v_j & \sim \mathcal{N}(\theta_j, \lambda^{-1}_v I_k)
    \label{ctr_latent_vectors}
 \end{split}
 \end{align}

 where $\theta_j$ is the topic component for the article. Furthermore,

 \begin{align}
    \mathbb{E}[r_{ij} | u^T_i, \theta_j, \epsilon_j]
    = u^T_i(\theta_j + \epsilon_j)
 \end{align}

 The topic component is used to draw words from the article. For each word $w_{jn}$ a latent topic $z_{jn}$ is drawn from the topic component. For every topic we associate a distribution of words described by the simplex $\beta_{z_{jn}}$.

 \begin{align}
     z_{jn} & \sim \text{Mult}(\theta_j) \\
     w_{jn} & \sim \text{Mult}(\beta_{z_{jn}})
 \end{align} 

 Lastly, we sample from eqn. \ref{reward_distr} to generate the reward associated with user $i$ and item $j$.

\begin{figure}[t]
    \includegraphics[width=0.3\textwidth]{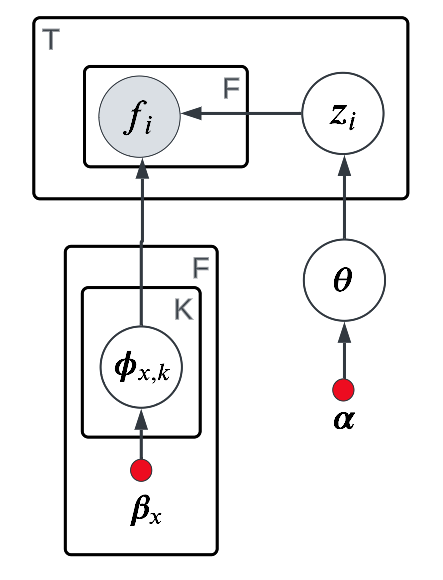}
    \centering
    \caption{The RatingMatch (RM) component of MV-ICTR, probabilistic diagram for generating topic components.}
    \label{RM_prob_diagram}
\end{figure}

 \subsection{Multi-view interactive collaborative topic regression}
The core idea idea behind CTR \cite{wang2011collaborative} is to estimate reward in a manner similar to PMF $r_{ij} = u_i^T v$, but with item vector $v_j = \theta_j + \epsilon_j$ represented as the sum of a topic component $\theta_j$ and a PMF offset component $\epsilon_j$. Arguably, we want want to express the user vector in a similar manner with $u_j = \theta^{(u)}_i + \epsilon^{(u)}_i$. Then,

\begin{align}
    r_{ij} = (\theta^{(u)}_i + \epsilon^{(u)}_i)^T(\theta^{(v)}_j + \epsilon^{(v)}_j)
    \label{goal_eqn}
\end{align}

We assume initially (i.e., before any ratings) that $\mathbb{E}[\theta_i^{(u)}] = \mathbb{E}[\theta_j^{(v)}] = 0$ and that $\theta^{(u)}_i \independent \epsilon^{(v)}_j$ and $\theta^{(v)}_j \independent \epsilon^{(u)}_i$, where $\independent$ denotes statistical independence, then

\begin{align}
    \mathbb{E}[r_{ij}] 
    \approx \mathbb{E}[\theta^{(u)T}_i \theta^{(v)}_j]
    \label{user_offset_eqn}
\end{align}

And herein lies the fundamental problem with the introduction of a user topic component $\theta^{(u)}_i$, it must be learned in congruence with $\theta^{(v)}_j$ such that eqn. \ref{user_offset_eqn} is sufficiently accurate for cold-start users and items. This is precisely the problem we address in this paper. If we are successful, we will have obtained a built-in method for addressing the cold-start problem in a way that treats users and items symmetrically.
 
 When user-item implicit or explicit rating pairs exist we are afforded the opportunity to associate user-item covariates. Specifically, user-specific features $f_i^{(u)}$ and item-specific features $f_j^{(v)}$ for any observed user-item pair $(i, j)$ can be concatenated:
 
 \begin{align}
     f_{ij} = [f_i^{(u)}; f_j^{(v)}] \in \mathcal{R}^{F}
 \end{align}
 
 where $F = F^{(u)} + F^{(v)} \in \mathcal{R}$ is the combined number of user-specific and item-specific features. MV-ICTR follows a similar line of thinking as \cite{wang2011collaborative} with ratings given by:

 \begin{align}
    r_{ij} \sim \mathcal{N}(u^T_i v_j, \sigma^2)
 \end{align}

 Now, we modify eqns. \ref{ctr_latent_vectors} to

  \begin{align}
  \begin{split}
    u_i & \sim \mathcal{N}(\chi^{(u)}_i, \lambda^{-1}_u I_K) \\
    v_j & \sim \mathcal{N}(\chi^{(v)}_j, \lambda^{-1}_v I_k)
    \label{mvictr_latent_vectors}
 \end{split}
 \end{align}

giving offset components to both user and items, where both are drawn from spherical normal distributions:
 \begin{align}
     \begin{split}
         \epsilon_i^{(u)} & \sim \mathcal{N}(0, \lambda^{-1}_u I_K) \\
         \epsilon_i^{(v)} & \sim \mathcal{N}(0, \lambda^{-1}_v I_K) \\
     \end{split}
 \end{align}

 we have finally latent user and item feature vectors, respectively: 
 \begin{align}
     \begin{split}
         u_i & = \chi^{(u)}_i + \epsilon_i^{(u)} \\
         v_i & = \chi^{(v)}_j + \epsilon_j^{(v)} \\
     \end{split}
 \end{align}
 
 Then, the conditional expectation of reward $r_{ij}$ is given by
  \begin{align}
    \mathbb{E}[r_{ij} | 
        \chi^{(u)}_i, \chi^{(v)}_j, \epsilon^{(u)}_i,  \epsilon^{(v)}_j]
        = (\chi^{(u)}_i + \epsilon_i^{(u)})^T(\chi^{(v)}_j + \epsilon^{(v)}_j)
        \label{cond_exp}
 \end{align}

where $\chi^{(u)}_i$ and $\chi^{(v)}_j$ give the conditional probabilities of user and item cluster assignments, respectively, and $\epsilon^{(u)}_i$ and $\epsilon^{(v)}_j$ their corresponding PMF offsets. Note that eqn. \ref{cond_exp} is identical in structure to eqn. \ref{goal_eqn}, which is what was desired. The vectors $\chi^{(u)}_i$ and $\chi^{(v)}_j$ are learned via the proposed RatingMatch (RM) procedure depicted in figure \ref{RM_prob_diagram}. The sampling procedure for RM is given below:

\medskip
\begin{enumerate}
    \item Draw the global distribution over clusters $\theta_i \sim \text{Dirichlet}(\alpha)$
    \item For each cluster $(k = 1,..., K)$ and for each feature $(x = 1,..., F)$
    \begin{enumerate}
        \item Draw $\phi_{x, k} \sim \text{Dirichlet}(\beta_x) \in \mathcal{R}^{V_x}$
    \end{enumerate}
    \item For each user-item pair $(i, j)$
    \begin{enumerate}
        \item Draw cluster assignment $z_{i} \sim \text{Mult}(\theta_i)$
        \item Draw user feature value $v_x \sim \text{Mult}(\phi_{x, z_{i}})$
    \end{enumerate}
\end{enumerate}
\medskip

To develop a procedure for learning $\chi^{(u)}_i$ and $\chi^{(v)}_j$ we first note that the expectations for Dirichlet variables $\theta^{(u)}_k$ and $\phi^{(u)}_{i, k}$ are given by:
 \begin{align}
    \mathbb{E}[\theta_k] & = \frac{n_k + \alpha}{\sum_{k=1}^K (n_k + \alpha)} 
    \label{global_assignment_exp} \\
    \mathbb{E}[\phi_{i, k}] & = 
    \prod_{x=1}^F \frac{
                n_{kf_{xv}} + \beta_x
            }{
                \sum_{v=1}^{V_x} (n_{kf_{xv}} + \beta_x)
            }
 \end{align}

  where $n_{kf_{xv}}$ is the number of times the feature $x$ belonging to ontology $f$ with value $v$ associated with datapoint $i$ is assigned to the $k^{th}$ cluster. Note that $n_{kf_{xv}}$ can be generalized as the corresponding \textit{sum of ratings points}. With this generalization, we therefore allow ratings $r_{i,j} \in [0, \infty)$. 
  
  It is helpful to think of each feature $x$ as having their own matrix of counts of size $K \times V_x$. In general, a feature $x$ can have multiple values $v = [v_1, ..., v_R]$, such as a movie with multiple genres. Then, we take

  \begin{align}
      n_{kf_{xv}} = \sum_{r = 1}^{R} n_{kf_{xv_r}}
  \end{align}
  
  Each feature is equipped with its own ontology, such as movie genres, user occupations, or skills associated with jobs. The in or out-of-sample probability of assignment for $z^{(u)}_m = k$ for user $m$ with user-specific covariates $f^{(u)}_m$ is therefore given by:

 \begin{align}
 \begin{split}
    \chi^{(u)}_{m, k} & \equiv
    P(z = k | f_m^{(u)}) \\
    & = P(z = k) \prod_{x \in F^{(u)}}P(f_{m, xv}^{(u)} | z = k) \\
    & \propto \mathbb{E}[\theta_k] 
        \cdot 
        \mathbb{E}[\phi^{(u)}_{m, k}]
 \end{split}
    \label{topic_components}
 \end{align}

 Where,
\begin{align}
    \mathbb{E}[\phi^{(u)}_{m, k}] \equiv
    \prod_{x \in F^{(u)}} \frac{
                n_{kf_{xv}^{(u)}} + \beta_x
            }{
                \sum_{v=1}^{V_x} (n_{kf_{m, xv}^{(u)}} + \beta_x)
            }
\end{align}

  Note that $\mathbb{E}[\phi^{(u)}_{m, k}]$ is defined the same as eqn. \ref{global_assignment_exp} but with the product being taken over user-associated features only. A similar equation exists for items, with the $n^{th}$ item's topic component given by $\chi^{(v)}_{n, k} \equiv P(z = k | f_n^{(v)})$. Eqn. \ref{topic_components} tells us that the user or item latent components can be obtained by simply ``plugging-in" the associated feature values. For training we use collapsed Gibbs sampler similar to that derived in \cite{maurya2017bayesian} and \cite{griffiths2004finding}. The latter found that Gibbs was faster to convergence than either expectation propagation (EP) or variational inference (VI) in learning latent Dirichlet allocation (LDA) \cite{blei2003latent} components. The conditional probability of cluster assignment is given below:

 \begin{align}
    P(z^{(u)}_i = k | z^{(u)}_{-i})
        \propto
        (n_{k, -i} + \alpha)
        \prod_{x=1}^F \frac{
                n_{kf_{xv}, -i} + \beta_x
            }{
                \sum_{v=1}^{V_x} (n_{kf_{xv}, -i} + \beta_x)
            }
 \end{align}

 where $n_{kf_{xv}, -i}$ is defined the same as $n_{kf_{xv}}$ but with the $i^{th}$ datapoint removed. A similar equation can be written for items. Therefore, when running the collapsed Gibbs procedure it is okay to train on all non-zero user-item ratings, where ratings $r_{i,j} \in [0, \infty)$. 
 
 Furthermore, note that there are numerous ways we could contrive to generate user and item offset parameters $\epsilon^{(u)}_i$ and $\epsilon^{(v)}_j$, other than PMF, for instance ICTR \cite{wang2018online}. However, proceeding with PMF we can compute the posterior distribution for the user matrix $U$, where each row of $U$ is a user latent vector $u_i$. Letting, $\delta_{ij} = \{ (i, j) : \text{user } i \text{ rated item } j\}$ we have,
 
\begin{align*}
    P(U | R, V; \sigma^2, \sigma_u^2, \sigma_v^2) 
    \propto P(U) P(R | U, V) \\
    \propto \prod_{i=1}^M \mathcal{N}(u_i | \epsilon^{(u)}_i, \sigma_u^2)
    \prod_{j\in\delta_{ij}} \mathcal{N}(r_{ij} | u_i^T v_j , \sigma^2) \\
\end{align*}

Taking the logarithm of the result gives:
\begin{align*}
        & \sum_{i=1}^M 
            \frac{-1}{2\sigma^2} \left[
                \frac{\sigma^2}{\sigma_u^2}
                (u_i - \epsilon^{(u)}_i)^T(u_i - \epsilon^{(u)}_i)
        + \sum_{j\in\delta_{ij}}
            (r_{ij} - u_i^T v_j)^T (r_{ij} - u_i^T v_j)
        \right] \\
        = & \sum_{i=1}^M 
            \frac{-1}{2\sigma^2} \left[
                u_i^T \left(
                    \sum_{j\in\delta_{ij}} v_j v_j^T
                    + \frac{\sigma^2}{\sigma_u^2} I
                \right) u_i    
        - 2 u_i^T \left(
            \sum_{j\in\delta_{ij}}
            r_{ij} v_j
            + \epsilon_i
        \right)
        + \sum_{j\in\delta_{ij}}
            r_{ij}^2
            + \epsilon_i^T \epsilon_i
        \right] \\
    = & \sum_{i=1}^M 
        \log{P(u_i | \mu_i, \Sigma_i )}
\end{align*}

The posterior, like the prior, is a normal distribution. The update equations are therefore given by:

 \begin{align}
     \begin{split}
         & u_i = (D_i^T D_i + \lambda_u I_K)^{-1} (D_i^T r_{i} + \epsilon_i^{(u)}) \\
         & v_j = (B_j^T B_j + \lambda_v I_K)^{-1} (B_j^T r_{i} + \epsilon_i^{(v)}) \\
         & \Sigma^{(u)}_i = (D_i^T D_i + \lambda_u I_K)^{-1} \sigma^2 \\
         & \Sigma^{(u)}_i = (D_i^T D_i + \lambda_u I_K)^{-1} \sigma^2
     \end{split}
 \end{align}

where
\begin{align}
    \begin{split}
    D_i & = \sum_{\delta_{ij} = 1} v_i v_i^T \\
    B_i & = \sum_{\delta_{ij} = 1} u_i u_i^T \\
    \end{split}
\end{align}

are the feature or design matrices, and $\lambda_u = \sigma^2 / \sigma_u^2$ and $\lambda_v = \sigma^2 / \sigma_v^2$. These can be viewed as the posterior after observing ratings $R = [r_{ij}]$. The corresponding updates are therefore similar to those given by the usual matrix factorization equations \cite{wang2011collaborative}, \cite{zhao2013interactive} and which also appear in Contextual-Bandits (CB) \cite{li2010contextual}, and also reassemble the coefficient estimation for ridge regression.

Note that in CB the design matrices can be in general high-dimensional, and that inversion operations have complexity $\mathcal{O}(n^3)$ where $n$ is the number of dimensions. Algorithms which perform dimensionality-reduction or latent variable modeling therefore have potentially vastly improved computational times.

\section{Experimental results}

\begin{figure}[t]
    \includegraphics[width=0.9\textwidth]{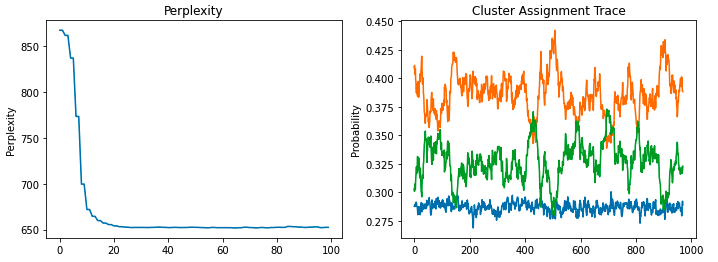}
    \centering
    \caption{MV-ICTR MovieLens 100K global cluster assignment trace with dimension of latent space given by $d=3$ over 1000 training dataset iterations, and associated perplexity over 100 training dataset iterations.}
\end{figure}

\begin{figure}[t]
    \includegraphics[width=\textwidth]{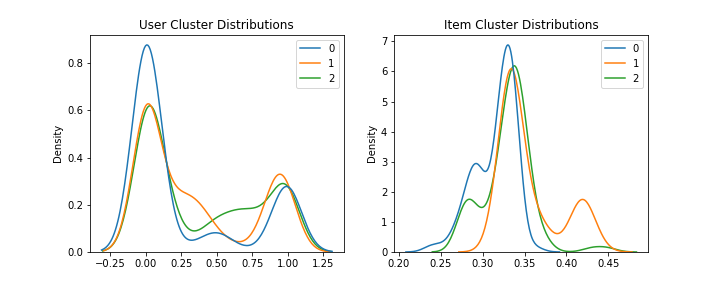}
    \centering
    \caption{MV-ICTR MovieLens 100K distributions of user and item clusters components $\chi^{(u)}_{n, k}$ and $\chi^{(v)}_{m, k}$, respectively, with the number of latent dimensions $d=3$, and $k\in[d]$. Taking the average of components over $n$ and $m$, we observed multi-modal peaks about 0 or 1 for users, and multi-modal peaks about 1/3 for items. The dimorphism in user and item distributions is likely due to the fact that no features are shared between user and items in this dataset.}
    \label{perplexity}
\end{figure}

\begin{figure}[t]
    \includegraphics[width=0.7\textwidth]{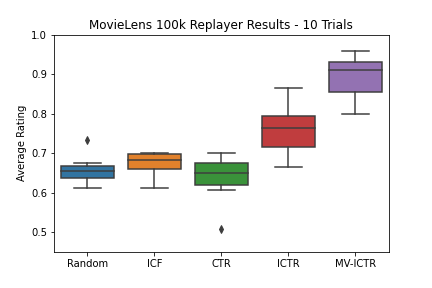}
    \centering
    \caption{Random, Thompson Sampling (TS), CTR, ICTR and MV-ICTR average rating of recommended items over 10 replayer trials. Rating performed significantly better than ICTR, the second highest performing model, with an average increase in performance of 13.5\% (two-sided t-test and p $< 0.01$).}
\end{figure}

To test the efficacy of our proposed algorithm we experimented on the Movie Lens 100K dataset, which was the only dataset we were able to locate with both ratings and covariates for all user and items. Regarding movies, we extracted two features ($F^{(v)} = 2$) release decade and genre. There were 8 unique decades ranging from 1920 to 1990 and 19 genres within the genre ontology. Regarding users, we used gender, occupation, and age measured in units of decades and rounded to the nearest integer. All features were treated as categorical. The ratings in MovieLens 100K were explicit and ranged from from 1 to 5. These were mapped to 1 if the rating was greater than or equal to 4 and 0 otherwise. We included only the 100 most popular (i.e., most rated) movies in our analysis, which helped to improve data efficiency in the offline evaluation experiments. After filtering, there remained 29.9K user-item ratings. 

Due the interactive nature of RatingMatch a rejection sampling approach was taken to evaluate the historical log of ratings data \cite{li2010contextual}. The approach allows us to acquire an unbiased estimate of the model performance but at the expense decreased data efficiency. The replayer method works by stepping through each rating event in the historical log and making a recommendation. If the recommended item does not correspond with the item from the historical log the event is simply discarded (rejected) and the algorithm proceeds to the next event. If the items do correspond, then the model parameters (if applicable) and rating score are updated. We therefore expect roughly 300 \textit{impressions}, or instances where the recommended item matches that in the historical log.

The data was ordered chronologically according to the event timestamps and split in half for training and testing. 97.7\% of datapoints in the test set are cold-start, meaning that either the user or the item rated were not found in the training set. If the online recommender included offset variables we trained these in the training set, and otherwise learned user-item rating-based components ``on-the-fly" in the test set. All reinforcement learning algorithms we're implemented with Thompson Sampling (TS), and in all cases where applicable, we chose the latent dimension of the design matrix or matrices to be $d=3$. We tested five different algorithms:

\textbf{Random}: Arm recommendations were randomly generated according to a uniform distribution. 

\textbf{Interactive Collaborative Filtering (ICF)}: ICF was performed with a Thompson Sampling policy. 

\textbf{Collaborative Topic Regression (CTR)}: Vanilla LDA was trained on the training set and used as the item-offsets in the test set. Following \cite{wang2011collaborative} we chose user and item precision parameters $\lambda_u = 0.01 $ and $\lambda_v = 100$, respectively. User and item rating-based latent vectors were learned sequentially on the test set, and policy decisions made with TS. 

\textbf{Interactive Collaborative Filtering (ICTR)}: ICTR was directly applied to the test set with no pre-training on the training set with a TS policy for online inference. Following \cite{wang2018online} we chose the dimension of the latent user-item vectors to be $d=3$, and the number of particles $B=10$.

\textbf{MV-ICTR}: RM learned user-item feature dependent offset vectors on the training set via collapsed Gibbs sampling. The offsets were then combined to untrained rating-based user and item vectors with $\lambda_u = \lambda_v = 1$ and $\sigma=0.01$ and implemented with a TS policy on the test set. RM was trained for 1000 iterations over the training set (see Figure \ref{perplexity}).

 Due to the stochastic nature of the recommendations and the restrictive size of the test set we trialed each algorithm 10 times and reported the average rating for recommended items over all trials for each algorithm. Perplexity, the typical measure for convergence in language models, was used to measure RM convergence, and is given by

 \begin{align}
     \text{Perplexity} = \exp{
        \left(
            \sum_{i=1}^T \log{p(f_i)}
        \right)
     }
 \end{align}

where,
\begin{align}
    \begin{split}
    p(f_i) & = \sum_{z = 1}^d p(f_i, z) \\
        & = \sum_{z = 1}^d p(f_i | z = k) p(z = k)
    \end{split}
\end{align}

where $f_i = [f_n^{(u)}; f_m^{(v)}]$ are the concatenated user and item features for the $i^{th}$ datapoint.
 \subsection{Results and discussion}
MV-ICTR was found to significantly increase the average rating on the test set by 13.5\% points over ICTR, the second best performing state-of-the-art algorithm tested in the application. We attribute the large jump in performance to the fact that the test set was 97.7\% composed of cold-start datapoints.

In summary, MV-ICTR combines RatingMatch, a multi-view clustering algorithm, with PMF for interactive recommendation. It saves computational time by reducing the dimension of the design matrix, and further saves on computation by learning the RM topic components offline. It was found to significantly outperform comparable leading state-of-art algorithms in the space of sequential or interactive bandit-based recommender systems, particularly on datasets with high percentages of cold-start users and items. It also generalizes well: RM is a multi-view clustering algorithm, capable of clustering user and items with overlapping, partially overlapping, or non-overlapping feature sets. RM also allows for out-of-sample predictions and for ratings assignments $r_{ij} \in {[0, \infty)}$, and because it is Bayesian, it is able to easily manage missing data.

MV-ICTR utilizes all of the strengths of RM and generalizes via PMF. Albeit, with enough rating data ICF should perform comparably, in real life applications user interaction data can be sparse and limiting, with high item turnover rates leading to high cold-start percentages. MV-ICTR solves the cold-start and personalization problem simultaneously, making it an optimal choice for applications such as article or job recommendation when user and item contextual information are available.

\subsection{Future work}
In future work, we would like to experiment with building a fully online topic model where both topic and rating components are updated after receiving feedback. We are also interesting in combining RM with ICTR for simultaneous user-item feature and item dependency modelling. Such a model could implement a particle filtering (PF) algorithm for online inference. We would also like to extend the research to different datasets, and to experiment with more bandit algorithms and different parameter values of the latent dimension, specifically comparing performance against computational efficiency.

\bibliographystyle{authordate1}
\bibliography{references}
 
\end{document}